\documentclass[journal]{IEEEtran}
\usepackage{cite}
\usepackage{graphicx}
\usepackage{amsmath}
\usepackage{times}
\usepackage{latexsym}
\usepackage{graphicx}
\usepackage{bm}
\usepackage{amssymb}
\usepackage[center]{caption2}
\usepackage{stfloats}
\usepackage{cases}
\usepackage{url}
\usepackage{array}
\usepackage{setspace}
\usepackage{fancyhdr}
\usepackage{citesort}
\usepackage{color}
\usepackage{eurosym}
\usepackage{subfigure}

\newtheorem{lemma}{Lemma}
\newtheorem{corollary}{Corollary}
\newtheorem{proposition}{Proposition}

\def\proof{\noindent\hspace{2em}{\itshape Proof: }}
\def\endproof{\hspace*{\fill}~$\square$\par\endtrivlist\unskip}

\begin{document}
\title{Wireless Information and Power Transfer with Full Duplex Relaying}
\author{Caijun Zhong,~\IEEEmembership{Senior Member,~IEEE}, Himal A. Suraweera,~\IEEEmembership{Member,~IEEE}, Gan Zheng,~\IEEEmembership{Senior Member,~IEEE}, Ioannis Krikidis,~\IEEEmembership{Senior Member,~IEEE}, and Zhaoyang Zhang~\IEEEmembership{Member,~IEEE},

\thanks{Caijun Zhong and Zhaoyang Zhang are with the Institute of Information and Communication Engineering, Zhejiang University, China. (email:caijunzhong@zju.edu.cn). Caijun Zhong is also with the National Mobile Communications Research Laboratory, Southeast University, Nanjing, 210018, China.}
\thanks{Himal A. Suraweera is with the Department of Electrical \& Electronic
Engineering, University of Peradeniya, Peradeniya 20400, Sri Lanka (email:
himal@ee.pdn.ac.lk)}
\thanks{G. Zheng is with the School of Computer Science \& Electronic Engineering, University of Essex, Wivenhoe Park, Colchester, CO4 3SQ, UK (e-mail: ganzheng@essex.ac.uk)}
\thanks{I. Krikidis is with the Department of Electrical \& Computer Engineering, University of Cyprus, Nicosia 1678, Cyprus (e-mail: krikidis.ioannis@ucy.ac.cy)}}

\markboth{IEEE TRANSACTIONS ON Communications,~Vol.~x, No.~xx,~xx~200x} {IEEE TRANSACTIONS ON COMMUNICATIONS,~Vol.~x,
No.~xx,~xx~200x}

\maketitle

\begin{abstract}
We consider a dual-hop full-duplex relaying system, where the energy constrained relay node is powered by radio frequency signals from the source using the time-switching architecture, both the amplify-and-forward and decode-and-forward relaying protocols are studied. Specifically, we provide an analytical characterization of the achievable throughput of three different communication modes, namely, instantaneous transmission, delay-constrained transmission, and delay tolerant transmission. In addition, the optimal time split is studied for different transmission modes. Our results reveal that, when the time split is optimized, the full-duplex relaying could substantially boost the system throughput compared to the conventional half-duplex relaying architecture for all three transmission modes. In addition, it is shown that the instantaneous transmission mode attains the highest throughput. However, compared to the delay-constrained transmission mode, the throughput gap is rather small. Unlike the instantaneous time split optimization which requires instantaneous channel state information, the optimal time split in the delay-constrained transmission mode depends only on the statistics of the channel, hence, is suitable for practical implementations.
\end{abstract}

\section{Introduction}
Conventional energy-constrained communication systems have a limited operational lifetime, and in order to maintain network connectivity, periodical battery replacement or recharging is performed, which is nevertheless costly, inconvenient and sometimes impossible. As such, energy harvesting, which scavenges energy from external natural resources such as solar, wind or vibration, has gained a great deal of interest, since it provides a cost-effective solution to prolong the lifetime of wireless communications systems. However, the amount of energy harvested from natural resources is random and highly depends on some uncontrollable factors such as the weather conditions, which makes reliable communication difficult. An interesting solution that overcomes the above limitation is to harvest energy from man-made radio frequency (RF) electromagnetic radiation (also known as wireless power transfer) \cite{W.Lumpkins,M.Pinuela}.

Since RF signals can carry both information and energy, there has been a tremendous upsurge of research activities in the area of simultaneous wireless information and power transfer (SWIPT). In the pioneering works on SWIPT by Varshney \cite{L.Varshney} and Grover \cite{P.Grover}, the fundamental tradeoff between the capacity and energy was studied. Later in \cite{R.Zhang}, practical architectures, i.e., time-switching and power-splitting, for SWIPT systems were proposed, and the optimal transmit covariance achieving the rate-energy region was derived. The extension of imperfect channel state information (CSI) at the transmitter was addressed in \cite{M.Tao}. More recently, sophisticated architectures improving the rate-energy region were proposed in \cite{L.Liu,X.Zhou}. In addition, the energy efficiency of OFDMA systems with SWIPT was studied in \cite{Derrick}, and the application of SWIPT in multiuser systems and cellular networks was considered in \cite{A.Fou,K.Huang}.

In parallel with the aforementioned works which mainly focus on the single hop scenario, employing intermittent relays to facilitate RF energy harvesting and information transfer has also drawn significant attention. The work in \cite{Metha} investigated the symbol error rate of relay selection in cooperative networks, where energy-constrained relay nodes with limited battery reserves rely on some external charging mechanism to assist the source-destination information transmission. In \cite{A.Nasir1}, the authors studied the throughput performance of an amplify-and-forward (AF) relaying system for both time-switching and power-splitting protocols, and later on, the same authors extended the analysis to the adaptive time-switching protocol in \cite{A.Nasir2}. The throughput of decode-and-forward relaying systems was investigated in \cite{A.Nasir3}, while the power allocation strategies for DF relaying system with multiple source-destination pairs was studied in \cite{Z.Ding1}. More recently, the performance of energy harvesting cooperative networks with randomly distributed users was studied in \cite{Z.Ding2,Z.Ding3,I.Krikidis2}. It is worth pointing out that all these works are limited to the half-duplex (HD) relaying mechanism, where the relay node can not receive and transmit data simultaneously in the same frequency band.

The HD architecture is widely adopted in traditional wireless relaying systems, because it can simplify the system design and implementation, it however incurs significant loss of spectrum efficiency. With the advance in antenna technology and signal processing capability, and in an effort to recover the spectral loss, full-duplex (FD) relaying, where the relay node receives and transmits simultaneously in the same frequency band, has received a lot of research interest (see references \cite{T.Riihonen1,T.Riihonen2,B.Day,D.Guo,I.Krikidis,H.Suraweera} and therein). However, to the best of the authors' knowledge, no works have considered the application of FD relaying in RF energy harvesting systems.

Motivated by this, we focus on a source-relay-destination dual-hop scenario where the relay is powered via RF energy harvesting, and investigate the effect of FD transmission on the system throughput in a RF energy harvesting relaying system. As for the FD relay, we consider the separate antenna configuration, i.e., the relay is equipped with two antennas, one for information transmission and one for information reception. In addition, the time-switching protocol is adopted.\footnote{Please note, with two antennas, it is possible to realize another type of FD operation, i.e., simultaneous energy harvesting and information transmission. However, in this paper, we adopt the time-switching protocol, i.e., energy harvesting and information transmission occur in disjoint time period, and the FD operation is only applied during the information transmission phase.} We study the throughput of both AF and DF relaying protocols, and characterize the fundamental trade-off between energy harvesting time and communication time. In particular, according to how the time split is optimized, three different communication modes are investigated, namely, instantaneous optimization based transmission which will be referred to as the instantaneous transmission hereafter, delay-constrained transmission, and delay tolerant transmission. In order to demonstrate the effect of the FD relaying architecture, the HD relaying architecture is also investigated. The main conclusion of this paper is that FD relaying is an attractive and promising solution to enhance the throughput of RF energy harvesting relay systems.

The main contributions of the paper are summarized as follows:
\begin{itemize}
\item We propose the idea of employing both antennas at the relay to scavenge energy during the energy harvesting period, and demonstrate that, with optimal time split, the dual antenna case always outperforms the single antenna case. However, the performance gap gradually diminishes when the source transmit power is sufficiently large.
\item For both AF and DF relaying systems, we present analytical expressions for the system throughput in all three different transmission modes. Specifically, analytical expressions for the outage probability are derived in the delay-constrained mode, while analytical expressions for the achievable rate are derived in the delay tolerant mode.
\item For AF relaying systems, we obtain the optimal time split through numerical calculation for all three transmission modes. For DF relaying systems, closed-form expressions for the optimal time split are presented for the instantaneous transmission mode, while the optimal time split for the delay-constrained and delay tolerant transmission modes is obtained numerically.
\item Comparing the AF and DF relaying protocols, our results show that the DF protocol always yields better throughput performance.
\item Our findings show that, with the optimal time split, the instantaneous transmission mode achieves a higher throughput than the delay tolerant transmission mode. However, the throughput benefit is not significant. Hence, taking into account of the optimization overhead required for the instantaneous transmission mode, i.e., each channel realization, the delay tolerant transmission mode is preferred in practice since it does not require the instantaneous CSI and the optimization is performed once over a relatively long period of time.
\item Comparing the FD and HD relaying architectures, our results demonstrate that, for a properly optimized systems, FD relaying  could substantially boost the throughput performance.
\end{itemize}

The rest of paper is organized as follows: Section II introduces the FD relaying systems, Section III deals with the case where the relay adopts a single antenna for energy harvesting, while Section IV focuses on the case where the relay adopts dual antennas for energy harvesting. Section V presents the corresponding performance of HD systems. Numerical results are presented in Section VI. Finally, Section VII summarizes the key findings of the paper.

\section{FD Relaying}
\begin{figure}[htb!]
\centering
\includegraphics[scale=0.5]{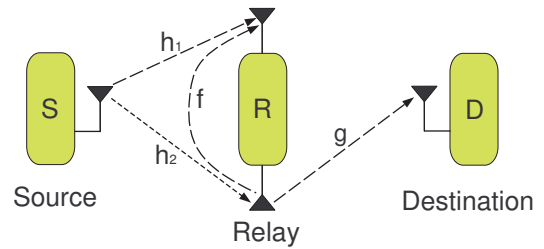}
\caption{System model, where S, R and D denote the source, relay and destination, respectively.}\label{fig:fig0}
\end{figure}

Let us consider a dual-hop FD relaying system illustrated in Fig. \ref{fig:fig0}, where the source sends information to the destination with the help of an intermediate relay. We assume that the source to destination link does not exist. It is also assumed that the relay only has limited power supply, and relies on external charging through harvesting energy from the source transmission \cite{A.Nasir1}.

We adopt the time-sharing protocol proposed in \cite{A.Nasir1}, hence the whole communication process is divided into two phases, i.e., the energy harvesting phase and the information transmission phase. Let $T$ denote the block time of an entire communication period during which some amount of information is transmitted from the source to the destination, then the first $\alpha T$ amount of time is used for harvesting energy at the relay, while the remaining $(1-\alpha)T$ amount of time is used for information transmission.

\begin{figure}[htb!]
\centering
\includegraphics[scale=0.5]{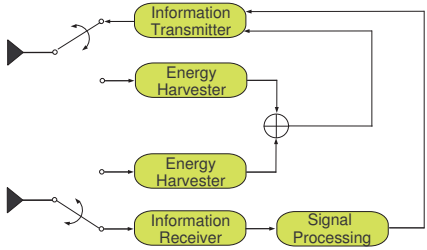}
\caption{Time-switching architecture.}\label{fig:fig01}
\end{figure}

To enable FD communication, the relay is equipped with two antennas, i.e., one for reception and one for transmission during the information transmission phase \cite{T.Riihonen1}. In addition, the two antennas can also be exploited during the energy harvesting phase, see Fig. \ref{fig:fig01}. Motivated by this, we consider two separate cases depending on the number of antennas used for energy harvesting as:
\begin{itemize}
\item[1.] Only the information receiving antenna is used to collect energy.
\item[2.] Both antennas are used to collect energy.
\end{itemize}
It is worth noting that Case 2 appears as a natural choice since it fully exploits the available hardware resources (antenna elements) to capture more energy. Nevertheless, due to its relatively easy implementation, Case 1 could be of practical interest in certain applications. In addition, as will be demonstrated in Section VI, Case 1 could achieve comparable performance as Case 2, especially in the high SNR regime. Therefore, it is also important to gain a deep understanding on the performance of Case 1. In the following, we give a detailed discussion on the signal models of both cases, the corresponding analysis will be presented in Section IV and Section V, respectively.

Let us begin with Case 1. During the energy harvesting phase, the received signal at the relay node can be expressed as
\begin{align}
y_r=\frac{h_1}{\sqrt{d_1^m}}x_e+n_r,
\end{align}
where $h_1$ is the channel coefficient, $d_1$ is the distance between the source and relay, while $m$ is the path loss exponent. In this work, we assume a normalized path loss model in order to show the path loss degradation effects on the system performance. In real-world scenarios, path loss significantly reduces the system performance and therefore potential scenarios are limited to near-field applications such as sensor \cite{H.Nishimoto,H.Visser} and wearable/body networks \cite{T.Sun}. $x_e$ is the energy symbol with ${\tt E}\{|x_e|^2\} = P_s$, where ${\tt E}\{x\}$ denotes the expectation operation. $n_r$ is the zero mean additive white Gaussian noise (AWGN) with variance $N_0$.

As in \cite{A.Nasir1}, we assume that the energy harvested during the energy harvesting phase is stored in a supercapacitor and then fully consumed by the relay to forward the source signal to the destination.\footnote{Please note, this is also known in the literature as harvest-use architecture \cite{A.Kaligh,S.Sude,I.Krikidis1} as opposed to the harvest-store-use architecture \cite{O.Ozel,F.Iannello}.} Hence, the relay transmit power can be computed as \cite{X.Zhou}\footnote{{Please note, the FD operation may require some extra energy for the mitigation of loopback interference. However, in this work, we do not consider such energy consumption. Hence, our model is particularly suitable for the scenario where loopback interference cancellation is based on passive suppression techniques such as antenna directionality \cite{E.Everett}.}}
\begin{align}\label{eqn:pr}
P_r = \frac{\eta P_s|h_{1}|^2\alpha T}{d_1^m(1-\alpha)T}=\frac{kP_s|h_{1}|^2}{d_1^m},
\end{align}
where $\eta$ is a constant and denotes the energy conversion efficiency and $k$ is defined as $k\triangleq\frac{\eta\alpha}{1-\alpha}$.

For Case 2, where both antennas are used to collect energy during the first phase, the received signal at the relay can be expressed as
\begin{align}
{\bf y}_r=\frac{1}{\sqrt{d_1^m}}{h_1 \choose h_2}x_e+{\bf n}_r,
\end{align}
where $h_2$ is the channel coefficient and ${\bf n}_r$ is the zero mean AWGN noise vector at the relay with ${\tt E}\{{\bf n}_r{\bf n}_r^{\dag}\} = N_0{\bf I}$, where ${\bf I}$ is the identity matrix. Similarly, the relay transmit power can be computed as
\begin{align}\label{eqn:pr}
P_r = \frac{\eta P_s(|h_{1}|^2+|h_2|^2)\alpha T}{d_1^m(1-\alpha)T}=\frac{kP_s(|h_{1}|^2+|h_2|^2)}{d_1^m}.
\end{align}

Now, let us look at the information transmission phase, the received signal at the relay can be expressed as
\begin{align}
y_r[i]=\frac{h_1}{\sqrt{d_1^m}}x_s[i]+fx_r[i]+n_r[i],
\end{align}
where $x_s[i]$ is the information symbol from the source at time slot $i$, and satisfies ${\tt E}\{|x_s[i]|^2\} = P_s$. $x_r[i]$ is the loopback interference due to full duplex relaying and satisfies ${\tt E}\{|x_r[i]|^2\} = P_r$, $f$ denotes the loopback interference channel, and $n_r[i]$ is the zero mean AWGN with variance $N_0$. Since the relay is aware of its own signal $x_r[i]$, it can apply interference cancellation methods to mitigate the loopback interference. Hence, the post-cancellation signal at the relay can be expressed as \cite{T.Riihonen1}
\begin{align}
\hat{y}_r[i]=\frac{h_1}{\sqrt{d_1^m}}x_s[i]+\hat{f}\hat{x}_r[i]+n_r[i],
\end{align}
where ${\tt E}\{|\hat{x}_r[i]|^2\}=P_r$ and $\hat{f}$ models the residual loopback interference channel due to imperfect cancellation \cite{T.Riihonen1}.

In this paper, we consider both AF and DF relaying protocols. With the AF protocol, the relay amplifies the input signal by a factor $\beta$ which is given by \cite{T.Riihonen1}
\begin{align}
\beta^2 = \frac{P_r}{|h_1|^2P_s/d_1^m+|\hat{f}|^2P_r+N_0}.
\end{align}
With the DF protocol, the relay first decodes the original signal and then regenerates the signal. Hence, the transmit signal of the relay can be expressed as \cite{T.Riihonen1}
\begin{align}
x_r[i] = \left\{\begin{array}{cc}
                    \beta \hat{y}_r[i-\tau], & \mbox{with AF},\\
                    \sqrt{\frac{P_r}{P_s}}x[i-\tau],&\mbox{with DF},
                    \end{array}\right.
\end{align}
where $\tau$ accounts for the time delay caused by relay processing. Therefore, the received signal at the destination can be expressed as
\begin{align}
y_d[i]=\frac{g}{\sqrt{d_2^m}}x_r[i]+n_d[i],
\end{align}
where $g$ is the channel coefficient, $d_2$ is the distance between the relay and destination, and $n_d[i]$ is the zero mean AWGN with variance $N_0$.


Fading assumptions: $|h_1|^2$ and $|h_2|^2$ are independent and identically distributed (i.i.d.) exponential random variables with mean $\lambda_{s}$, $|g|^2$  is exponentially distributed with mean $\lambda_{d}$, and $|\hat{f}|^2$ is exponentially distributed with mean $\lambda_{r}$, which is a key parameter related to the strength of the loopback interference.

\section{Energy harvesting with single antenna}
In this section, we study the throughput performance of FD relaying with RF energy harvesting using a single antenna and investigate the optimal time allocation strategy. Specifically, we consider three different communication modes, i.e., instantaneous transmission, delay-constrained transmission, and delay tolerant transmission. For mathematical tractability, we focus on the loopback interference dominated scenario which is of practical interest \cite{D.Guo}.

At this point, it is important to clarify the CSI requirement of three different transmission modes. In general, it is assumed that CSI is available at the destination node. For the instantaneous transmission mode, the optimal time split is updated for each channel realization, which should be computed by a centralized entity having access to the global instantaneous CSI. On the other hand, for the delay constrained transmission and delay tolerant transmission modes, only the channel statistics are required to compute the optimal time split. In addition, the optimal time split is updated only if the channel statistics change, i.e., the distance and the fading parameter etc, which in general varies much slower. Hence, the overhead associated with the CSI acquisition of these two modes is substantially smaller compared with the instantaneous transmission mode.

\subsection{Instantaneous Transmission}
We now focus on the instantaneous throughput, and examine the optimal $\alpha$ for both the AF and DF protocol. Let us start with the AF protocol.

\subsubsection{AF}
With the AF protocol, the end-to-end SINR can be expressed as
\begin{align}\label{eqn:2}
\gamma_{\sf AF}&=\frac{\frac{P_s|h_{1}|^2}{P_r|\hat{f}|^2d_1^md_2^m}P_r|g|^2}{\frac{N_0P_s|h_{1}|^2}{P_r|\hat{f}|^2d_1^m}+\frac{P_r|g|^2}{d_2^m}+N_0},
\end{align}
hence, the instantaneous throughput is given by
\begin{multline}
R_{\sf IA}(\alpha)=(1-\alpha)\times\\
\log_2\left(1+\frac{\frac{P_s|h_{1}|^2|g|^2}{|\hat{f}|^2d_1^md_2^m}}{\frac{N_0(1-\alpha)}{\eta\alpha|\hat{f}|^2}+\frac{\eta\alpha}{1-\alpha}\frac{P_s|h_{1}|^2|g|^2}{d_1^md_2^m}+N_0}\right).
\end{multline}
The optimal $\alpha$ could be obtained by solving the following optimization problem
\begin{align}
&\alpha^* = \mbox{arg} \max_{\alpha}R_{\sf IA}(\alpha)\notag\\
&\mbox{subject to }0<\alpha<1.
\end{align}
Given the fact that $R_{\sf IA}(\alpha)$ is a concave function of $\alpha$, the optimal value $\alpha^*$ can be obtained by solving the equation
$\frac{d R_{\sf IA}(\alpha)}{d\alpha} = 0$. However, due to the complexity of the involved expression, a closed-from solution is not possible. Instead in this work, the optimal value $\alpha^*$ is numerically evaluated using the build-in function ``NSlove'' of Mathematica.

\subsubsection{DF}
The end-to-end SINR of the DF protocol can be expressed as
\begin{align}
\gamma_{\sf DF}= \min\left\{\frac{1}{k|\hat{f}|^2},\frac{kP_s|h_{1}|^2|g|^2}{d_1^md_2^mN_0}\right\},
\end{align}
and hence, the instantaneous throughput is given by
\begin{multline}
R_{\sf ID}(\alpha) =(1-\alpha)\times \\ \log_2\left(1+\min\left\{\frac{1-\alpha}{\eta\alpha|\hat{f}|^2},\frac{\eta\alpha}{1-\alpha}\frac{P_s|h_{1}|^2|g|^2}{d_1^md_2^mN_0}\right\}\right).
\end{multline}
The optimal $\alpha$ could be obtained by solving the following optimization
\begin{align}
&\alpha^* = \mbox{arg} \max_{\alpha}R_{\sf ID}(\alpha)\notag\\
&\mbox{subject to }0<\alpha<1.
\end{align}
The above optimization could solved analytically, and we have the following key result:
\begin{proposition}\label{prop:1}
The optimal $\alpha^*$ is given by
\begin{align}\label{alpha:1}
\alpha^* =\left\{\begin{array}{cc} \frac{e^{ W\left(\frac{c_2-1}{e}\right)+1}-1}{{c_2-1+e^{ W\left(\frac{c_2-1}{e}\right)+1}}}, & \mbox{ if }e^{ W\left(\frac{c_2-1}{e}\right)+1}<\frac{c_2}{\alpha_0}\\
\frac{1}{1+\alpha_0}, & \mbox{otherwise},
\end{array}
\right.
\end{align}
where $c_1 = \eta|\hat{f}|^2$, $c_2=\frac{\eta P_s|h_1|^2|g_1|^2}{d_1^md_2^mN_0}$, $\alpha_0^2 = c_1c_2$ and $W(x)$ is the Lambert W function, where $W(x)$
is the solution of $W \exp(W) = x$.
\end{proposition}
\proof See Appendix \ref{appendix:prop:1}. \endproof

\subsection{Delay-constrained Transmission}
For delay-constrained transmission, the source transmits at a constant rate $R_c$, which may subject to outage due to the random fading of the wireless channel. Hence, the average throughput can be computed as \cite{A.Nasir1}
\begin{equation}\label{throughput:dc}
R_{\sf DL}(\alpha)=(1-P_{\sf out})R_c(1-\alpha),
\end{equation}
where $P_{\sf out}$ is the outage probability. Hence, the optimal time portion $\alpha$ can be obtained from
\begin{equation}\label{eqn:optalpha1}
\alpha^* = \arg\max_{\alpha}  R_{\sf DL}(\alpha).
\end{equation}
Therefore, the remaining key task is to characterize the exact outage probability of the system.

\subsubsection{AF}
For the AF protocol, we have the following key result:
\begin{proposition}\label{prop:2}
For the AF protocol, the outage probability of the system can be expressed as
\begin{align}\label{eqn:afcdf}
&P_{\sf out}^{\sf AF} =1-\int_0^{\frac{1}{k\gamma_{\sf th}}}2\sqrt{\frac{d_1^md_2^m}{\lambda_{s}\lambda_{d}}\frac{\frac{\gamma_{\sf th}N_0}{k}+\gamma_{\sf th}N_0y}{P_s-kP_s\gamma_{\sf th}y}}\times\notag\\
&K_1\left(2\sqrt{\frac{d_1^md_2^m}{\lambda_{s}\lambda_{d}}\frac{\frac{\gamma_{\sf th}N_0}{k}+\gamma_{\sf th}N_0y}{P_s-kP_s\gamma_{\sf th}y}}\right)\frac{1}{\lambda_{r}}e^{-\frac{y}{\lambda_{r}}}dy,
\end{align}
where $K_n(x)$ is the $n$-th order modified Bessel function of the second kind \cite[Eq. (8.432.1)]{Table}.
\end{proposition}
\proof See Appendix \ref{appendix:prop:2}.\endproof


\subsubsection{DF}
%
%

For the DF protocol, we have the following key result:
\begin{proposition}\label{prop:3}
For the DF protocol, the outage probability of the system can be expressed as
\begin{multline}\label{eqn:dfcdf}
P_{\sf out}^{\sf DF}=1-\left(1-e^{-\frac{1}{k\lambda_{r}\gamma_{\sf th}}}\right)\\\left(2\sqrt{\frac{d_1^md_2^mN_0\gamma_{\sf th}}{kP_s\lambda_{s}\lambda_{d}}}K_1\left(2\sqrt{\frac{d_1^md_2^mN_0\gamma_{\sf th}}{kP_s\lambda_{s}\lambda_{d}}}\right)\right).
\end{multline}
\end{proposition}
\proof The end-to-end SINR can be expressed as
\begin{align}
\gamma_{\sf DF}= \min\left\{\frac{1}{ky},\frac{kP_sx}{d_1^md_2^mN_0}\right\},
\end{align}
where $x=|h_{1}|^2|g|^2$ and $y=|\hat{f}|^2$. Now, applying the fact that random variables $x$ and $y$ are independent, the desired result follows after some algebraic manipulations.\endproof

Given the outage expressions in (\ref{eqn:afcdf}) and (\ref{eqn:dfcdf}), the optimization problem in (\ref{eqn:optalpha1}) does not admit a closed-form solution. However, the optimal $\alpha$ is efficiently solved via numerical calculation.

%

\subsection{Delay Tolerant Transmission}
In the delay tolerant scenario, the source transmits at any constant rate upper bounded by the ergodic capacity. Since the codeword length is sufficiently large compared to the block time, the codeword could experience all possible realizations of the channel. As such, the ergodic capacity becomes an appropriate measure. Please note, in the current work, the residual loopback interference is treated as Gaussian noise. Hence, the throughput of the system is given by
\begin{align}
R_{\sf DT} = (1-\alpha)R_{\sf E},
\end{align}
where $R_{\sf E}$ is the achievable rate of the system. Hence, the optimal time portion $\alpha$ can be obtained by solving
\begin{equation}\label{eqn:optalpha2}
\alpha^* = \arg\max_{\alpha}  R_{\sf DL}(\alpha).
\end{equation}
Therefore, the remaining key task is to characterize the exact ergodic capacity of the system, which we now study.

\subsubsection{AF}
For the AF protocol, we have the following key result:
\begin{proposition}\label{prop:4}
For the AF protocol, the ergodic capacity can be computed as
\begin{align}
R_{\sf E}^{\sf AF} &= \frac{G_{4,2}^{1,4}\left(\frac{kP_s\lambda_{s}\lambda_{d}}{N_0d_1^md_2^m}\left|_{1,0}^{0,0,1,1}\right.\right)}{\ln 2}+\notag\\
&\frac{1}{\lambda_r\ln 2}\int_0^{\infty}G_{4,2}^{1,4}\left(\frac{k^2P_s\lambda_{s}\lambda_{d}y}{N_0d_1^md_2^m(1+ky)}\left|_{1,0}^{0,0,1,1}\right.\right)e^{-\frac{y}{\lambda_r}}dy
,\end{align}
where $G_{m,n}^{p,q}(x)$ is the Meijer G-function \cite[Eq. (9.301)]{Table}.
\end{proposition}
\proof See Appendix \ref{appendix:prop:4}.\endproof

Since the above expression involves an integral, which in general does not admit a closed-form solution, hence, it is not amenable to further processing. Motivated by this, we now present the following tight upper bound.
\begin{corollary}
The ergodic capacity of the AF protocol is upper bounded by
\begin{align}
R_{\sf up}^{\sf AF} &= \frac{1}{\ln 2}e^{\frac{1}{k\lambda_{r}}}E_1\left(\frac{1}{k\lambda_{r}}\right)+\frac{G_{4,2}^{1,4}\left(\frac{kP_s\lambda_{s}\lambda_{d}}{N_0d_1^md_2^m}\left|_{1,0}^{0,0,1,1}\right.\right)}{\ln 2}\notag\\
&-\log_2\left(1+ke^{\psi(1)+\ln \lambda_{r}}+\frac{k^2P_s}{N_0d_1^md_2^m}e^{3\psi(1)+\ln \lambda_{r}\lambda_{s}\lambda_{d}}\right),
\end{align}
where $E_n(x)$ is the exponential integral function \cite[Eq. (5.1.4)]{handbook}, and $\psi(x)$ is the digamma function \cite[Eq. (8.360.1)]{Table}.
\end{corollary}
\proof
The proof relies on the fact that $f(x,y)=\log_2(1+e^x+e^y)$ is a convex function with respect to $x$ and $y$. Hence, we have
\begin{multline}
{\tt E}\left\{\log_2\left(1+ky+\frac{k^2P_sxy}{N_0d_1^md_2^m}\right)\right\}\geq\\
 \log_2\left(1+ke^{{\tt E}\left\{\ln y\right\}}+\frac{k^2P_s}{N_0d_1^md_2^m}e^{{\tt E}\left\{\ln x\right\}+{\tt E}\left\{\ln y\right\}}\right).
\end{multline}
Hence, the remaining task is to compute ${\tt E}\left\{\ln x\right\}$ and ${\tt E}\left\{\ln y\right\}$. With the help of the integration relationship \cite[Eq. (4.352.1)]{Table}, we have
\begin{align}
{\tt E}\left\{\ln y\right\} = \frac{1}{\lambda_{r}}\int_0^{\infty}\ln y e^{-\frac{y}{\lambda_{r}}}dy = \psi(1)+\ln \lambda_{r}.
\end{align}
Similarly, we get ${\tt E}\left\{\ln x\right\} =  2\psi(1)+\ln \lambda_{s}\lambda_{d}$, which completes the proof.

\subsubsection{DF}
The achievable rate of the DF protocol is given by the proposition below:
\begin{proposition}
The ergodic capacity of the DF protocol can be expressed as
\begin{multline}
R_{\sf DT}^{\sf DF}
=\frac{1}{\ln 2}\times\\
\int_0^{\infty}\frac{\left(1-e^{-\frac{1}{k\lambda_{r}t}}\right)\left(2\sqrt{\frac{d_1^md_2^mN_0t}{kP_s\lambda_{s}\lambda_{d}}}K_1\left(2\sqrt{\frac{d_1^md_2^mN_0t}{kP_s\lambda_{s}\lambda_{d}}}\right)\right)}{1+t}dt\notag
\end{multline}
\end{proposition}
\proof
Utilizing the cumulative distribution function (c.d.f.) of the end-to-end SINR given in (\ref{eqn:dfcdf}), the desired result can be obtained.
\endproof

\section{Energy harvesting with dual antennas}
In this section, we consider the scenario where both antennas at the relay are used during the energy harvesting phase. Similarly, we first characterize the achievable throughput in three different transmission scenarios, and then investigate the optimal time split.

\subsection{Instantaneous Transmission}
\subsubsection{AF}
Recall that when both antennas are used to collect energy during the first phase, the harvested energy is given by
\begin{align}
P_r=\frac{kP_s(|h_{1}|^2+|h_2|^2)}{d_1^m}.
\end{align}
Hence, for the interference dominated scenario, the end-to-end SINR can be written as
\begin{align}\label{eqn:af2snr}
\gamma_{\sf AF}&=\frac{\frac{P_s|h_{1}|^2|g|^2}{|\hat{f}|^2d_1^md_2^m}}{\frac{N_0|h_1|^2(1-\alpha)}{\eta\alpha(|h_1|^2+|h_2|^2)|\hat{f}|^2}+\frac{\eta\alpha P_s(|h_{1}|^2+|h_2|^2)|g|^2}{(1-\alpha)d_1^md_2^m}+N_0},
\end{align}
%
and as a result, the instantaneous throughput is given by
\begin{align}
&R_{\sf IA}(\alpha) =(1-\alpha)\log_2\left(1+\gamma_{\sf AF}\right).
\end{align}
The optimal $\alpha$ could be obtained by solving the following optimization problem
\begin{align}
&\alpha^* = \mbox{arg} \max_{\alpha}R_{\sf IA}(\alpha)\notag\\
&\mbox{subject to }0<\alpha<1.
\end{align}
Again, due to the complexity of the involved expression, a closed-form solution of the optimal $\alpha$ is not possible. Instead, the optimal value $\alpha^*$ is numerically evaluated using the build-in function ``NSlove'' of the Mathematica software.

\subsubsection{DF}
For the DF protocol, the end-to-end SINR can be expressed as
\begin{align}
\gamma_{\sf DF}= \min\left\{\frac{|h_1|^2}{k|\hat{f}|^2(|h_1|^2+|h_2|^2)},\frac{kP_s(|h_{1}|^2+|h_2|^2)|g|^2}{d_1^md_2^mN_0}\right\},
\end{align}
hence, the instantaneous throughput is given by
\begin{align}
&R_{\sf ID}(\alpha) = (1-\alpha)\log_2\left(1+\gamma_{\sf DF}\right).
\end{align}
Having characterized the end-to-end SINR, the optimal $\alpha$ can be obtained using the proposition below.
\begin{proposition}
The optimal $\alpha^*$ is given by
\begin{align}\label{alpha:2}
\alpha^* =\left\{\begin{array}{cc} \frac{e^{ W\left(\frac{c_4-1}{e}\right)+1}-1}{{c_4-1+e^{ W\left(\frac{c_4-1}{e}\right)+1}}}, & \mbox{ if }e^{ W\left(\frac{c_4-1}{e}\right)+1}<\frac{c_4}{\alpha_3}\\
\frac{1}{1+\alpha_3}, & \mbox{otherwise},
\end{array}
\right.
\end{align}
where $c_3 = \frac{\eta |\hat{f}|^2(|h_1|^2+|h_2|^2)}{|h_1|^2}$, $c_4=\frac{\eta P_s(|h_1|^2+|h_2|^2)|g|^2}{d_1^md_2^mN_0}$, and $\alpha_3^2 = c_3c_4$.
\end{proposition}
\proof
The proof is similar to that of Proposition \ref{prop:1} and is omitted.
\endproof

\subsection{Delay-constrained Transmission}
We now consider the delay-constrained scenario, and we start with the AF protocol.
\subsubsection{AF}
From (\ref{eqn:af2snr}),
we are ready to study the outage probability of the system.
\begin{proposition}\label{theorem:af2}
The outage probability of the system with the AF protocol can be expressed as
\begin{align}
&P_{\sf out}^{\sf AF} =1-\int_{k\gamma_{\sf th}}^{\infty}\frac{2N_0d_1^md_2^m\left(\frac{\gamma_{\sf th}t}{k}+\gamma_{\sf th}\right)}{\lambda_{s}\lambda_{d}(tP_s-kP_s\gamma_{\sf th})}\times\notag\\
&K_2\left(2\sqrt{\frac{N_0d_1^md_2^m\left(\frac{\gamma_{\sf th}t}{k}+\gamma_{\sf th}\right)}{\lambda_{s}\lambda_{d}(tP_s-kP_s\gamma_{\sf th})}}\right)\left(\lambda_{r}-\left(\lambda_{r}+\frac{1}{t}\right)e^{-\frac{1}{\lambda_{r}t}}\right)dt
\end{align}
\end{proposition}
\proof See Appendix \ref{appendix:theorem:af2}.\endproof

%
\subsubsection{DF}
For the DF protocol,
we have the following key result.

\begin{proposition} \label{prop:8}
The outage probability of the system with the DF protocol is given by
\begin{multline}\label{eqn:cdfull-duplexf1}
P_{\sf out}^{\sf DF} = 1-\left(1-\lambda_{r}\gamma_{\sf th}k\left(1-e^{-\frac{1}{\lambda_{r}\gamma_{\sf th}k}}\right)\right)\frac{2\gamma_{\sf th}N_0d_1^md_2^m}{kP_s\lambda_{s}\lambda_{d}}\\K_2\left(2\sqrt{\frac{\gamma_{\sf th}N_0d_1^md_2^m}{kP_s\lambda_{s}\lambda_{d}}}\right).
\end{multline}
\end{proposition}
\proof See Appendix \ref{app:prop:8}.\endproof

\subsection{Delay Tolerant Performance}
\subsubsection{AF}
The achievable rate of the system with the AF protocol is given in the following proposition.
\begin{proposition}\label{prop:9}
The ergodic capacity of AF relaying systems can be computed as
\begin{align}
&R_{\sf E}^{\sf AF} = \frac{1}{\ln 2}G_{3,1}^{1,3}\left(\left.ka\lambda_s\lambda_d\right|_{1}^{-1,1,1}\right)-\frac{1}{\ln 2}\times\\
&\int_0^{\infty}G_{3,1}^{1,3}\left(\left.\frac{ka\lambda_s\lambda_d}{1+t/k}\right|_{1}^{-1,1,1}\right)\left(\lambda_{r}-\left(\lambda_{r}+\frac{1}{t}\right)e^{-\frac{1}{\lambda_{r}t}}\right)dt.
\end{align}
\end{proposition}
\proof See Appendix \ref{app:prop:9}.\endproof

Alternatively, we can use the following closed-form upper bound.
\begin{corollary}\label{coro:1}
The ergodic capacity of the AF protocol is upper bounded by
\begin{multline}
R_{\sf up}^{\sf AF} = \frac{e^{\frac{1}{k\lambda_r}}}{\ln 2}\left(E_1\left(\frac{1}{k\lambda_r}\right)+k\lambda_r\sum_{k=1}^2E_k\left(\frac{1}{k\lambda_r}\right)
\right)\\-\frac{1}{\ln 2}\left(1+\psi(1)+\ln k\lambda_r+k\lambda_r(\psi(2)+\ln k \lambda_r)\right)\\+\frac{G_{4,2}^{1,4}\left(\frac{kP_s\lambda_{s}\lambda_{d}}{N_0d_1^md_2^m}\left|_{1,0}^{0,0,1,1}\right.\right)}{\ln 2}
\\-\log_2\left(1+\frac{1}{k\lambda_r}e^{-1-\psi(1)}+ka\lambda_s\lambda_de^{2\psi(1)}\right).
\end{multline}
\end{corollary}
\proof See Appendix \ref{app:coro:1}.\endproof

\subsubsection{DF}
The achievable rate of the DF protocol is given in the following proposition.
\begin{proposition}
The ergodic capacity of the DF protocol can be expressed as
\begin{multline}
R_{\sf DT}^{\sf DF}
=\frac{1}{\ln 2}\\\int_0^{\infty}\frac{\left(1-\lambda_{r}kt\left(1-e^{-\frac{1}{\lambda_{r}kt}}\right)\right)\frac{2tN_0d_1^md_2^m}{kP_s\lambda_{s}\lambda_{d}}K_2\left(2\sqrt{\frac{tN_0d_1^md_2^m}{kP_s\lambda_{s}\lambda_{d}}}\right)}{1+t}dt.\notag
\end{multline}
\end{proposition}
\proof
The desired result follows easily by invoking (\ref{eqn:cdfull-duplexf1}).
\endproof
\section{HD Relaying}
In this section, we present the corresponding results for the HD relaying system, which serves as a benchmark for performance comparison. Similar to the FD relaying system, we consider three different transmission modes. Some of the results have been derived in \cite{A.Nasir1}, nevertheless, we present them here to make the current work self-contained.

We begin with a brief description of the HD relaying signal model. The energy harvesting phase is the same as the FD relaying system. For the information transmission phase, the remaining $(1-\alpha)$ portion of block time is equally partitioned into two time slots. In the first time slot, the source transmits to the relay, and the received signal at the relay is given by
\begin{align}
y_r=\frac{h_1}{\sqrt{d_1^m}}x_s+n_r.
\end{align}
With the AF protocol, the relay amplifies the input signal by a factor $\beta$ which is given by
\begin{align}
\beta^2 = \frac{P_r}{|h_1|^2P_s/d_1^m+N_0},
\end{align}
where $P_r = \frac{2kP_s|h_1|^2}{d_1^m}$. Then, the transmit signal of the relay can be expressed as
\begin{align}
x_r = \left\{\begin{array}{cc}
                    \beta {y}_r, & \mbox{with AF},\\
                    \sqrt{\frac{P_r}{P_s}}x_s,&\mbox{with DF}.
                    \end{array}\right.
\end{align}
Therefore, the received signal at the destination can be expressed as
\begin{align}
y_d=\frac{g}{\sqrt{d_2^m}}x_r+n_d.
\end{align}

\subsection{Instantaneous Transmission}
\subsubsection{AF}
The end-to-end SNR of HD relaying can be computed as
\begin{align}
\gamma_{\sf AF} = \frac{\frac{P_sP_r|g|^2|h_1|^2}{d_1^md_2^mN_0^2}}{\frac{|g|^2P_r}{d_2^mN_0}+\frac{|h_1|^2P_s}{N_0d_1^m}+1}.
\end{align}
Hence, the instantaneous throughput is given by
\begin{align}
R_{\sf IA}(\alpha) = \frac{1-\alpha}{2}\log_2\left(1+ \frac{\frac{P_sP_r|g|^2|h_1|^2}{d_1^md_2^mN_0^2}}{\frac{|g|^2P_r}{d_2^mN_0}+\frac{|h_1|^2P_s}{N_0d_1^m}+1}\right).
\end{align}


\subsubsection{DF}
The end-to-end SNR of HD DF relaying scheme can be computed as
\begin{align}
\gamma_{\sf DF} = \min\left(\frac{P_s|h_1|^2}{N_0d_1^m},\frac{2kP_s|h_1|^2|g|^2}{N_0d_1^md_2^m}\right).
\end{align}
The instantaneous throughput is given by
\begin{align}
R_{\sf ID}(\alpha) = \frac{1-\alpha}{2}\log_2\left(1+ \min\left(\frac{P_s|h_1|^2}{N_0d_1^m},\frac{2\eta \alpha P_s|h_1|^2|g|^2}{N_0d_1^md_2^m(1-\alpha)}\right)\right).
\end{align}
\begin{proposition}\label{prop:6}
The optimal $\alpha^*$ is given by
\begin{align}\label{alpha:3}
\alpha^* =\left\{\begin{array}{cc} \frac{e^{ W\left(\frac{c_5-1}{e}\right)+1}-1}{{c_5-1+e^{ W\left(\frac{c_5-1}{e}\right)+1}}}, & \mbox{ if }e^{ W\left(\frac{c_5-1}{e}\right)+1}<\frac{c_5d_2^m}{2\eta|g|^2}+1,\\
\frac{1}{1+\frac{2\eta|g|^2}{d_2^m}} & \mbox{otherwise},
\end{array}
\right.
\end{align}
where $c_5= \frac{2\eta P_s|h_1|^2|g|^2}{N_0d_1^md_2^m}$.
\end{proposition}
\proof The proof is similar to that of Proposition \ref{prop:1} and is omitted.
\endproof

\subsection{Delay-constrained Transmission}
For the HD relaying system, the throughput is given by
\begin{equation}
R_{\sf DL}^{\sf HD}(\alpha)=\frac{1}{2}(1-P_{\sf out})R(1-\alpha),
\end{equation}
where $1/2$ accounts for the HD constraint.

\subsubsection{AF}
This case has been studied in \cite{A.Nasir1}, and we present the result here for the sake of completeness.
\begin{lemma}
For the HD mode, the outage probability of the AF protocol can be computed as
\begin{align}
P_{\sf out}^{\sf AF} = 1-\frac{1}{\lambda_{s}}\int_{\frac{\gamma_{\sf th}}{P_s}}^{\infty}e^{-\frac{x}{\lambda_{s}}-\frac{\gamma_{\sf th}\left(\frac{xP_s}{N_0d_1^m}+1\right)}{\lambda_{d}\left(\frac{2kP_s^2x^2}{N_0d_1^md_2^m}-\frac{2k\gamma_{\sf th}P_sx}{N_0d_1^md_2^m}\right)}}dx.
\end{align}
\end{lemma}

\subsubsection{DF}
For HD DF relaying systems, the end-to-end SNR is given by
\begin{align}
\gamma_{\sf DF} = \min\left(\frac{P_s|h_1|^2}{N_0d_1^m},\frac{2kP_s|h_1|^2|g|^2}{N_0d_1^md_2^m}\right).
\end{align}
\begin{proposition}\label{prop:10}
The outage probability of HD DF relaying systems is given by
\begin{align}\label{prop:101}
P_{\sf out}^{\sf DF}=1-e^{-\frac{N_0d_1^m\gamma_{\sf th}}{\lambda_{s}P_s}-\frac{d_1^m}{2\lambda_{d}k}}-\frac{1}{\lambda_{d}}\int_0^{\frac{d_1^m}{2k}}e^{-\frac{N_0d_1^md_2^m\gamma_{\sf th}}{2k\lambda_{s}P_sx}}e^{-\frac{x}{\lambda_{d}}}dx.
\end{align}
\end{proposition}
\proof
Starting from definition of the outage probability, we have
\begin{align}
P_{\sf out}&= {\sf Pr}\left(\frac{P_s|h_1|^2}{N_0d_1^m}\min\left(1,\frac{2k|g|^2}{d_2^m}\right)<\gamma_{\sf th}\right).
\end{align}
Now, conditioned on $y = \min\left(1,\frac{2k|g|^2}{d_2^m}\right)$, we get $P_{\sf out}=1-e^{-\frac{N_0d_1^m\gamma_{\sf th}}{\lambda_{s}P_sy}}$. Averaging over $y$ yields the desired result.
\endproof

\subsection{Delay tolerant performance}
For HD relaying system, the throughput is given by
\begin{equation}
R_{\sf DT}^{\sf HD}(\alpha)=\frac{1}{2}(1-\alpha)R_{\sf E}.
\end{equation}
\subsubsection{AF}
The AF protocol has been considered in \cite{A.Nasir1}, where the exact achievable rate expression involves double integral. Here we present an alternative expression which requires a single integral.
\begin{corollary}
The ergodic capacity of the AF protocol is upper bounded by
\begin{align}
&R_{\sf E}^{\sf AF} = \frac{G_{4,2}^{1,4}\left(\frac{2kP_s\lambda_{s}\lambda_{d}}{N_0d_1^md_2^m}\left|_{1,0}^{0,0,1,1}\right.\right)}{\ln 2}-\frac{e^{\frac{d_2^m}{2k\lambda_d}}}{\lambda_s\ln2}\times\notag\\
&\int_0^{\infty}e^{\frac{d_1^md_2^mN_0}{2k\lambda_d P_sx}}E_1\left(\frac{d_1^md_2^mN_0+P_sd_2^mx}{2k\lambda_d P_sx}\right)e^{-\frac{x}{\lambda_s}}dx.
\end{align}
\end{corollary}
\proof
The achievable rate of the system can be alternatively computed by
\begin{align}
&R_{\sf E}^{\sf AF} = {\tt E}\left\{\log_2\left(1+\frac{2kP_s|h_1|^2|g|^2}{d_1^md_2^mN_0}\right)\right\}-\notag\\
&{\tt E}\left\{\log_2\left(1+\frac{2kP_s|h_1|^2}{d_1^md_2^mN_0+P_s|h_1|^2d_2^m}|g|^2\right)\right\}.
\end{align}
Since the first item has been computed in (\ref{eqn:keyhole}) in Appendix \ref{appendix:prop:4}, we focus on the second item. Now, conditioned on $|h_1|^2$, we have
\begin{multline}
{\tt E}\left\{\log_2\left(1+\frac{2kP_s|h_1|^2}{d_1^md_2^mN_0+P_s|h_1|^2d_2^m}|g|^2\right)\right\}=\\\frac{1}{\ln2}e^{\frac{d_1^md_2^mN_0+P_s|h_1|^2d_2^m}{\lambda_d2kP_s|h_1|^2}}E_1\left(\frac{d_1^md_2^mN_0+P_s|h_1|^2d_2^m}{\lambda_d2kP_s|h_1|^2}\right).
\end{multline}
Then, averaging over $|h_1|^2$ yields the desired result.
\endproof

\subsubsection{DF}

\begin{proposition}
The achievable rate of the DF protocol can be expressed as
\begin{multline}
R_{\sf E}^{\sf DF}
=\frac{\lambda_sP_s}{N_0d_1^m}e^{\frac{N_0d_1^m}{\lambda_sP_s}-\frac{d_1^m}{2k\lambda_{d}}}E_1\left(\frac{N_0d_1^m}{\lambda_sP_s}\right)+\\\frac{1}{\lambda_d\ln 2 }\int_0^{\infty}\frac{1}{1+t}dt\int_0^{\frac{d_1^m}{2k}}e^{-\frac{N_0d_1^md_2^mt}{2k\lambda_{s}P_sx}}e^{-\frac{x}{\lambda_{d}}}dx.
\end{multline}
\end{proposition}
\proof
Starting from the c.d.f. of the end-to-end SNR given in (\ref{prop:101}),
the desired result can be obtained after some algebraic manipulations.
\endproof
%

\section{Numerical Results}
In this section, we present numerical simulation results to validate out analytical expressions developed in the previous section, and investigate the impact of key system parameters on the throughput of the system. Unless otherwise stated, we set the source transmission rate as $R_c = 3 \mbox{ bps/Hz}$, hence the outage SINR threshold is given by $\gamma_{\sf th} = 2^{R_c}-1 = 7$. The energy harvesting efficiency is set to be $\eta = 0.4$, while the path loss exponent is set to be $m=3$. The distances $d_1$ and $d_2$ are normalized to unit value. Also, we set $\lambda_{s} = 1$, $\lambda_{r} = 0.1$ and $\lambda_{d} = 1$, unless otherwise specified.

\begin{figure}[ht]
  \centering
  \subfigure[Outage probability: $\alpha=0.2$]{\label{fig:1a}\includegraphics[width=0.45\textwidth]{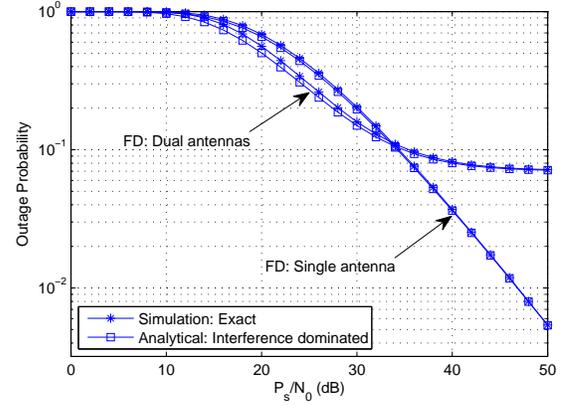}}
  \hspace{0.2in}
  \subfigure[Achievable rate: $\alpha=0.1$]{\label{fig:1b}\includegraphics[width=0.45\textwidth]{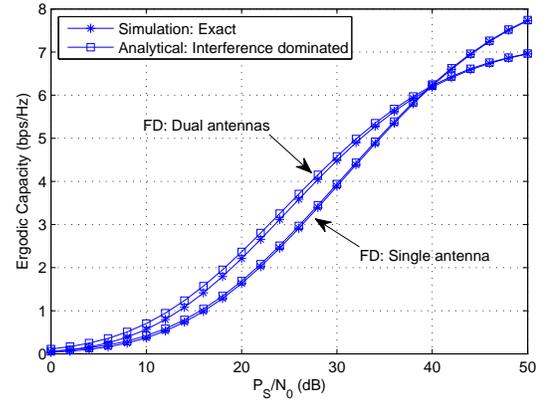}}
  \caption{Validation of the interference dominated assumption.}
  \label{fig:fig1}
\end{figure}

For mathematical tractability, the assumption of interference-limited relay is adopted, i.e., the noise at the relay is ignored. We now validate this assumption in Fig. \ref{fig:fig1}. As we can readily observe, for both the outage probability and achievable rate, the interference-limited curves provide a very good approximation to the exact performance. We also notice that the two antenna case outperforms the single antenna case in the low SNR regime, while the opposite is true in the high SNR regime. When the transmit power is small, the two antenna case can harvest more energy to facilitate the information transmission. However, when the transmit power is large, excessive amount of energy is collected, which is actually detrimental since it causes strong loopback interference, which degrades the system's performance. Please note, this phenomenon is due to the fact that a fixed $\alpha$ is adopted regardless of the transmit power. In general, superior energy harvesting capability is always beneficial. One can simply tune $\alpha$ to achieve better performance. As will be illustrated in the following figures, the dual antenna case always outperforms the single antenna case with optimized $\alpha$.


\begin{figure}[ht]
  \centering
  \subfigure[DF]{\label{fig:2a}\includegraphics[width=0.45\textwidth]{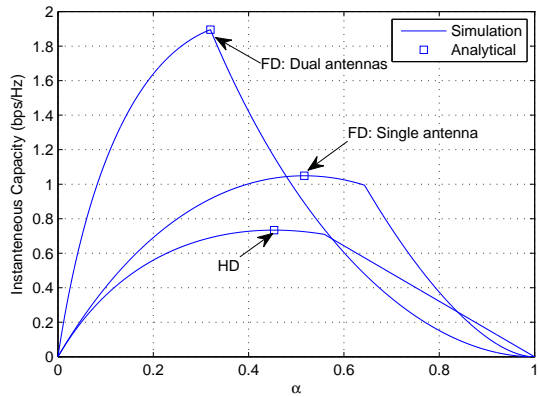}}
  \hspace{0.2in}
  \subfigure[AF]{\label{fig:2b}\includegraphics[width=0.45\textwidth]{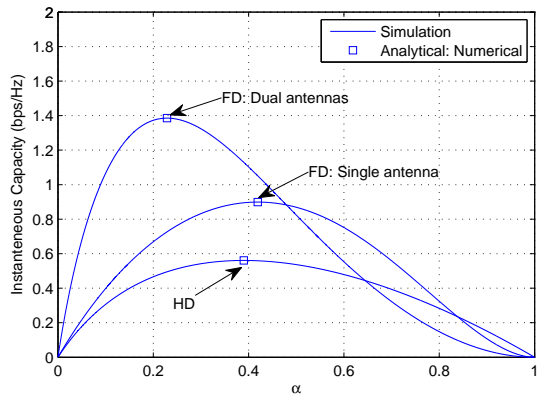}}
  \caption{Instantaneous capacity versus $\alpha$: $|h_1|^2=0.83$, $|h_2|^2=2.35$, $|g|^2=0.986$, $|f|^2=0.235$, $P_s/N_0=10 \mbox{ dB}$.}
  \label{fig:fig2}
\end{figure}

Fig. \ref{fig:fig2} deals with the impact of optimal time split $\alpha$ on the instantaneous capacity. We focus on a single time frame with the following setting: $|h_1|^2=0.83$, $|h_2|^2=2.35$, $|g|^2=0.986$, $|f|^2=0.235$, $P_s/N_0=10 \mbox{ dB}$. As can be readily observed, the analytical solutions given in (\ref{alpha:1}), (\ref{alpha:2}), and (\ref{alpha:3}) are in exact agreement with the simulation results. Interestingly, unlike the phenomenon observed \cite{I.Krikidis1}, that the optimal $\alpha$ for the HD case is smaller than the FD single antenna case, the optimal $\alpha$ for the FD two antenna case is smaller than the HD case. This is intuitive since the additional antenna helps reduce the energy harvesting time. Similar conclusions can be drawn for the AF protocol as shown in Fig. \ref{fig:2b}. In addition, we see that the optimal $\alpha$ for the AF protocol tends to be smaller than that of the DF protocol.

\begin{figure}[ht]
  \centering
  \subfigure[FD: Single antenna]{\label{fig:3a}\includegraphics[width=0.45\textwidth]{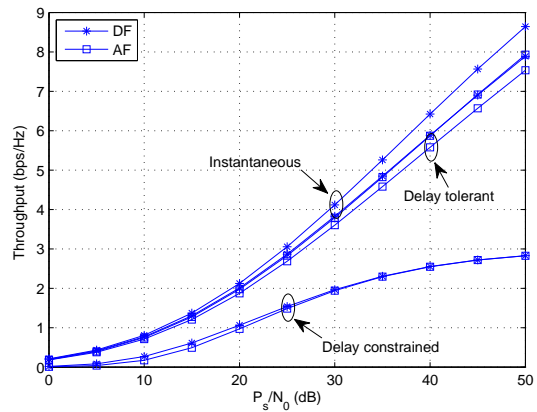}}
  \hspace{0.2in}
  \subfigure[FD: Dual antennas]{\label{fig:3b}\includegraphics[width=0.45\textwidth]{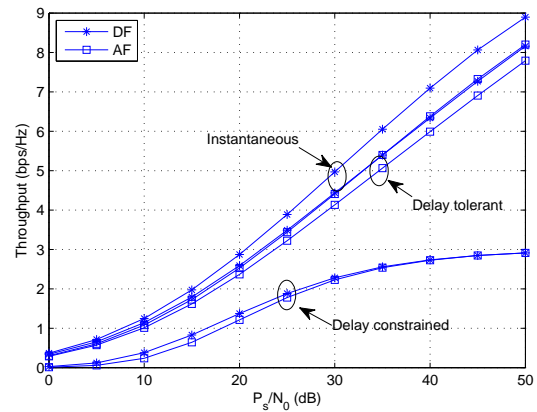}}
  \caption{Throughput comparison with optimized $\alpha$.}
  \label{fig:fig3}
\end{figure}

Fig. \ref{fig:fig3} illustrates the throughput of different scenarios with the optimal $\alpha$. We see from both Fig. \ref{fig:3a} and Fig. \ref{fig:3b}, that the throughput with instantaneous optimization is always the largest. Nevertheless, the throughput advantage appears to be limited when compared with the delay tolerant scenario. This indicates that the stochastic optimization approach may be preferred since it requires much less overhead. Also, as expected, the
throughput of the delay constrained scenario is upper bounded by the constant transmission rate $R_c = 3\mbox{ bps/Hz}$. Moreover, we observe that the DF scheme outperforms the AF scheme slightly in all scenarios. Now, comparing the single antenna case and dual antenna case, we find out that the dual antenna case always attains a higher throughput, especially in the low SNRs. Take the throughput of AF scheme with instantaneous optimization as an example, when $\frac{P_s}{N_0}=20 \mbox{ dB}$, the throughput of the single antenna case is about $2 \mbox{ bps/Hz}$, while the corresponding throughput is $2.5 \mbox{ bps/Hz}$. This is intuitive since at the low SNRs, the relay is energy constrained, hence the additional antenna could harvest more energy to facilitate the information transmission. As the transmit power increases, a single antenna can harvest sufficient amount of energy, hence, the advantage of having additional antenna diminishes gradually.

\begin{figure}[ht]
  \centering
  \subfigure[$\frac{P_s}{N_0} = 20\mbox{ dB}$]{\label{fig:4a}\includegraphics[width=0.45\textwidth]{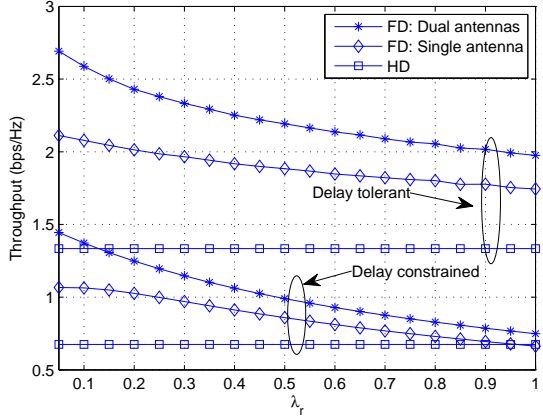}}
  \hspace{0.2in}
  \subfigure[$\frac{P_s}{N_0} = 50\mbox{ dB}$]{\label{fig:4b}\includegraphics[width=0.45\textwidth]{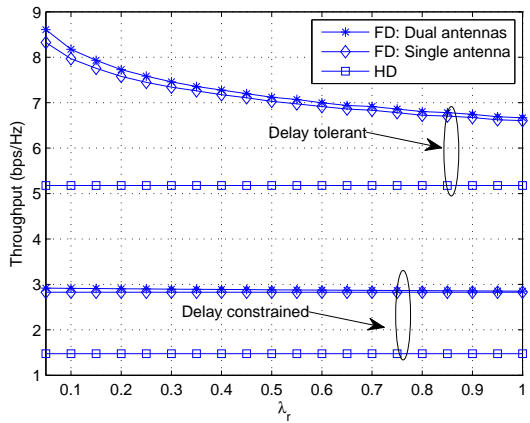}}
  \caption{Impact of $\lambda_r$ on the system throughput with DF protocol and optimized $\alpha$.}
  \label{fig:fig4}
\end{figure}

Fig. \ref{fig:fig4} illustrates the impact of $\lambda_r$ on the throughput of the DF relaying system with optimized $\alpha$. We observe that, when $\lambda_r$ increases, the throughput decreases, which is rather expected since a large $\lambda_r$ implies a stronger residual loopback interference. Also, the throughput loss is more pronounced for small values of $\lambda_r$, where a slight increase in $\lambda_r$ could significantly reduce the achievable throughput. In addition, we see that, as the source transmit power increases, the throughput difference of the two FD schemes becomes insignificant. This is due to the fact that, with high source transmit power, a single antenna could collect sufficient energy for information transmission, hence the advantage of employing an additional antenna to harvest energy is significantly reduced. As such, both FD schemes achieve a similar throughput. Another interesting observation is that with high source transmit power, the throughput of the delay constrained systems is insensitive to the level of loopback interference, i.e., the value of $\lambda_r$. This is also intuitive, since with high source transmit power, given a fixed transmission rate $R_c=3\mbox{ bps/Hz}$, the outage probability is close to zero, and varying $\lambda_r$ only makes a rather small change to the outage probability with optimized $\alpha$. Moreover, with high source transmit power, for the delay constrained scenario, the FD schemes almost achieve the maximum throughput of $3\mbox{ bps/Hz}$, which nearly doubles the maximum throughput of the HD scheme. This is because in such a scenario, the outage probability of both the FD and HD schemes approaches zero, hence, the penalty of the HD constraint is more significant. However, for the delay tolerant scenario, the throughput advantage of FD schemes significantly diminishes. This is due to the effect of residual loopback interference, which reduces the achievable rate of the FD schemes.

\begin{figure}[htb!]
\centering
\includegraphics[scale=0.6]{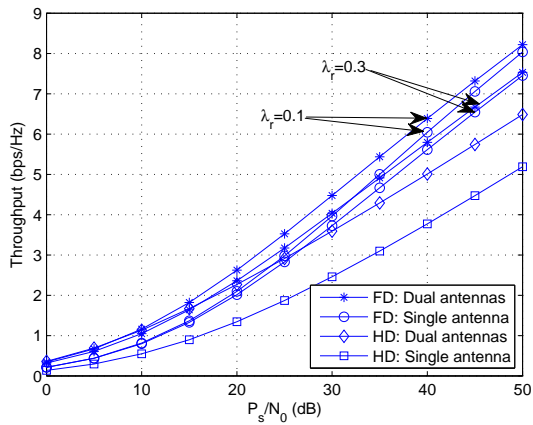}
\caption{Throughput comparison of delay tolerant systems with DF protocol and optimized $\alpha$: FD vs. HD.}\label{fig:fig5}
\end{figure}

Fig. \ref{fig:fig5} compares the throughput of delay tolerant systems with the DF protocol and optimized $\alpha$. In particular, the HD scheme with two antennas is also included. In such a system, it is assumed that the relay applies maximum ratio combining during the information receiving stage, and employs maximum ratio transmission during the information forwarding stage. As can be readily observed, the HD dual antenna case always outperforms the HD single antenna case, while it outperforms the FD single antenna case in the low SNR regime, i.e., $P_s/N_0\leq 20 \mbox{ dB}$. Now, compared to the FD dual antenna case, we see that the HD dual antenna case attains a similar throughput in the low SNRs, and exhibits an inferior performance in the high SNRs. Moreover, the strength of the loopback interference is a key factor determining to what extent the superiority of the FD dual antenna case holds. For instance, with weak loopback interference, i.e., $\lambda_r=0.1$, the FD dual antenna case outperforms the HD dual antenna case when $P_s/N_0\geq 10 \mbox{ dB}$; with strong loopback interference, i.e., $\lambda_r=0.3$, the FD dual antenna case outperforms the HD dual antenna case when $P_s/N_0\geq 20 \mbox{ dB}$. Finally, it is worth pointing out that the implementation of the HD dual antenna case illustrated in Fig. \ref{fig:fig5} requires two pairs of RF chains as opposed to a single pair RF chain required for the FD dual antenna case. Since the cost of additional RF chain is quite significant, in this regard, the FD scheme is a cost-effective solution.

\begin{figure}[htb!]
\centering
\includegraphics[scale=0.6]{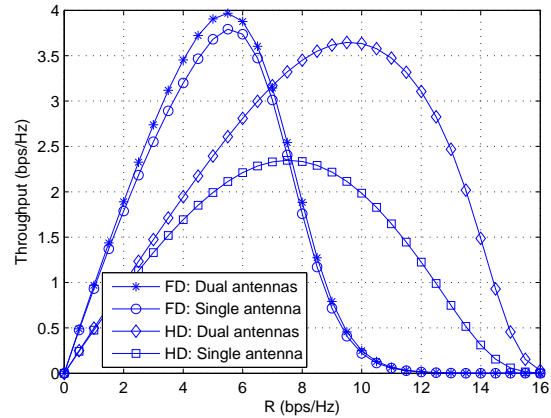}
\caption{Impact of transmission rate $R_c$ on the throughput of delay constrained systems with optimized $\alpha$.}\label{fig:fig6}
\end{figure}

Fig. \ref{fig:fig6} examines the impact of transmission rate $R_c$ on the throughput of delay constrained systems with optimized $\alpha$ when $P_s/N_0 = 40 \mbox{ dB}$. In all schemes, the throughput first increases with the transmission rate $R_c$, and then decreases when $R_c$ increases beyond a certain value. This is intuitive, since according to (\ref{throughput:dc}), when the transmission rate is small, the throughput is small; when the transmission rate is large, the outage probability increases significantly, which again degrades the throughput; hence, for a particular transmit power, there exists a unique transmission rate which yields maximum throughput. Interestingly, we see that the optimal transmission rate for the two FD schemes is the same, which is smaller than that of the HD schemes. This could be explained as follows: Because of the loopback interference in FD schemes and the doubled transmit power in HD scheme, the HD scheme achieves better outage performance than the FD schemes for any fixed transmission rate $R_c$. In other word, the HD schemes could tolerate a higher transmission rate $R_c$ without significant deterioration of the outage performance.

\section{Conclusion}
In this paper, we have studied the throughput of FD relaying in RF energy harvesting systems. Depending on the number of relay antennas used in the energy harvesting phase, two different cases were studied. For both the AF and DF protocols, analytical expressions were derived for the outage probability and ergodic capacity of the system. Based on which, the optimal time split was studied. It was demonstrated that employing both relay antennas for energy harvesting is always beneficial, and the throughput gain is most significant when the source transmit power is small. In addition, compared to the HD relaying architecture, our results indicate that FD relaying can substantially boost the system throughput with optimal time split. More importantly, a large portion of the potential gain offered by FD relaying can be realized using only the channel statistics without the need of instantaneous CSI, hence, it is a promising solution for implementing future RF energy harvesting cooperative systems.

\appendices
\section{Proof of Proposition \ref{prop:1}}\label{appendix:prop:1}
Define $\alpha_0^2 = c_1c_2$, where $c_1 = \eta|\hat{f}|^2$. We find it convenient to consider two separate regions, i.e., 1) $0<\alpha<\frac{1}{1+\alpha_0}$ and 2) $\frac{1}{1+\alpha_0}\leq \alpha<1$.

For the region 1), the throughput can be shown as
\begin{align}
C_{\sf ID}(\alpha) = (1-\alpha)\log_2\left(1+\frac{c_2\alpha}{1-\alpha}\right).
\end{align}
Taking the first derivative of $C_{\sf ID}(\alpha)$ with respect to $\alpha$ and setting $\frac{d C_{\sf ID}(\alpha)}{d\alpha}=0$, we have
\begin{align}
c_2+\frac{c_2\alpha}{1-\alpha}=\left(1+\frac{c_2\alpha}{1-\alpha}\right)\ln\left(1+\frac{c_2\alpha}{1-\alpha}\right).
\end{align}
Now, let $y = 1+\frac{c_2\alpha}{1-\alpha}$, the above equation can be written as
\begin{align}
c_2-1+y = y\ln y,
\end{align}
which, after some algebraic manipulations, can be expressed as
\begin{align}
\ln\left(\frac{y}{e}\right)e^{\ln\left(\frac{y}{e}\right)}=\frac{c_2-1}{e},
\end{align}
which is in the form of the standard definition of Lambert W function. Hence, we have
\begin{align}
\ln\left(\frac{y}{e}\right) = W\left(\frac{c_2-1}{e}\right).
\end{align}
After some simple algebraic manipulations, we get
\begin{align}
\alpha_1^* =\frac{e^{ W\left(\frac{c_2-1}{e}\right)+1}-1}{{c_2-1+e^{ W\left(\frac{c_2-1}{e}\right)+1}}}.
\end{align}

We now consider the second region $\frac{1}{1+\alpha_0}\leq \alpha<1$. In this case, we have
\begin{align}
C_{\sf ID}(\alpha) = (1-\alpha)\log_2\left(1+\frac{1-\alpha}{c_1\alpha}\right).
\end{align}
Taking the first derivative of $C_{\sf ID}(\alpha)$ with respect to $\alpha$ yields
\begin{align}
\frac{d C_{\sf ID}(\alpha)}{d\alpha} = -\log_2\left(1+\frac{1-\alpha}{c_1\alpha}\right)-\frac{1-\alpha}{c_1\alpha^2}\frac{1}{1+\frac{1-\alpha}{c_1\alpha}},
\end{align}
which is strictly smaller than zero. Hence, $ C_{\sf ID}(\alpha)$ is a decreasing function with respect to $\alpha$. Therefore, the optimal $\alpha$ is given by
\begin{align}
\alpha_2^* = \frac{1}{1+\alpha_0}.
\end{align}
Noticing that if $\alpha_1^*\in \left(0, \frac{1}{1+\alpha_0}\right)$, then $\alpha_1^*$ is optimal, otherwise, $\alpha_2^*$ is optimal.

\section{Proof of Proposition \ref{prop:2}}\label{appendix:prop:2}
Starting from the definition, the outage probability is given by
\begin{align}
P_{\sf out}^{\sf AF}= {\sf Pr}\left\{\gamma_{\sf AF}\leq \gamma_{\sf th}\right\},
\end{align}
where $\gamma_{\sf th} = 2^R-1$. Invoking the end-to-end SINR expression given in (\ref{eqn:2}), we have
\begin{align}
P_{\sf out}^{\sf AF}&={\sf Pr}\left\{\frac{\frac{P_s|h_{1}|^2}{P_r|\hat{f}|^2d_1^md_2^m}P_r|g|^2}{\frac{N_0P_s|h_{1}|^2}{P_r|\hat{f}|^2d_1^m}+\frac{P_r|g|^2}{d_2^m}+N_0}\leq \gamma_{\sf th}\right\}.
\end{align}
Define $x=|h_{1}|^2|g|^2$, $y=|\hat{f}|^2$, conditioned on $y$, the outage probability can be simplified as
\begin{equation}
P_{\sf out}^{\sf AF} = \left\{\begin{array}{cc} {\sf Pr}\left\{x\leq \frac{d_1^md_2^m\left(\frac{\gamma_{\sf th}N_0}{k}+\gamma_{\sf th}N_0y\right)}{P_s-kP_s\gamma_{\sf th}y}\right\},& y<\frac{1}{k\gamma_{\sf th}}\\
                1, & {y}>\frac{1}{k\gamma_{\sf th}},\\
                \end{array}
                \right.
\end{equation}
Utilizing the result in \cite{C.Zhong}, the c.d.f. of $x$ can be shown as
\begin{align}
F_x(x) = 1-2\sqrt{\frac{x}{\lambda_{s}\lambda_{d}}}K_1\left(2\sqrt{\frac{x}{\lambda_{s}\lambda_{d}}}\right).
\end{align}
To this end, averaging over $y$, the desired result can be obtained after some simple algebraic manipulations.

\section{Proof of Proposition \ref{prop:4}}\label{appendix:prop:4}
Define $x=|h_{1}|^2|g|^2$, $y=|\hat{f}|^2$, the ergodic capacity can be expressed as
\begin{align}
&R_{\sf E}^{\sf AF}={\tt E}\left\{\log_2\left(1+\frac{kP_sx}{N_0d_1^md_2^m}\right)\right\}\notag\\
&-{\tt E}\left\{\log_2\left(1+\frac{k^2P_sxy}{N_0d_1^md_2^m(1+{ky})}\right)\right\}
\end{align}
%
Noticing that the probability density function (p.d.f.) of $x$ is given by
\begin{align}
f(x) = \frac{2}{\lambda_{s}\lambda_{d}}K_0\left(2\sqrt{\frac{x}{\lambda_{s}\lambda_{d}}}\right),
\end{align}
the first item could be evaluated as
\begin{align}
&{\tt E}\left\{\log_2\left(1+\frac{kP_sx}{N_0d_1^md_2^m}\right)\right\} \notag\\
&=\frac{2}{\lambda_{s}\lambda_{d}\ln 2}\int_0^{\infty}\ln\left(1+\frac{kP_sx}{N_0d_1^md_2^m}\right)K_0\left(2\sqrt{\frac{x}{\lambda_{s}\lambda_{d}}}\right)dx\\
&=\frac{2}{\lambda_{s}\lambda_{r}\ln 2}\int_0^{\infty}G_{2,2}^{1,2}\left(\frac{kP_sx}{N_0d_1^md_2^m}\left|_{1,0}^{1,1}\right.\right)K_0\left(2\sqrt{\frac{x}{\lambda_{s}\lambda_{d}}}\right)dx
\end{align}
 and we have used the relationship \cite[Eq. (8.4.6.5)]{A.Prud}. Then, using the integral identity \cite[Eq. (7.821.3)]{Table},
we have
\begin{align}\label{eqn:keyhole}
{\tt E}\left\{\log_2\left(1+\frac{kP_sx}{N_0d_1^md_2^m}\right)\right\} &=\frac{G_{4,2}^{1,4}\left(\frac{kP_s\lambda_{s}\lambda_{d}}{N_0d_1^md_2^m}\left|_{1,0}^{0,0,1,1}\right.\right)}{\ln 2}.
\end{align}
Now, conditioned on $y$, the second item can be shown similarly as
\begin{multline}
{\tt E}\left\{\log_2\left(1+\frac{k^2P_sxy}{N_0d_1^md_2^m(1+{ky})}\right)\right\}=\\\frac{G_{4,2}^{1,4}\left(\frac{k^2P_s\lambda_{s}\lambda_{d}y}{N_0d_1^md_2^m(1+ky)}\left|_{1,0}^{0,0,1,1}\right.\right)}{\ln 2},
\end{multline}
the desired result then follows immediately.
\endproof

\section{Proof of Proposition \ref{theorem:af2}}\label{appendix:theorem:af2}
Let us define $U = \frac{X}{X+Y}$ and $V = X+Y$, with $X=|h_1|^2$ and $Y=|h_2|^2$, then the end-to-end SINR can be alternatively expressed as
\begin{align}
\gamma_{\sf AF}  &=\frac{\frac{U}{|\hat{f}|^2}\frac{P_sV|g|^2}{N_0d_1^md_2^m}}{\frac{U}{k|\hat{f}|^2}+\frac{kP_sV|g|^2}{N_0d_1^md_2^m}+1}.
\end{align}
To proceed, we first study the statistical property of random variables $U$ and $V$, and establish the independence of $U$ and $V$.

Due to the independence of $X$ and $Y$, the joint distribution of $X$ and $Y$ is given by
\begin{align}
f_{X,Y}(x,y) = \frac{1}{\lambda_s^2}e^{-\frac{x+y}{\lambda_s}}.
\end{align}
Also, it can be easily shown that\begin{align}
X = UV, \quad \quad\mbox{and}\quad\quad
Y = V(1-U).
\end{align}
Hence, the Jacobian of the transformation from $(U, V)$ back to $(X, Y)$ can be computed as
\begin{align}
\left|J\left(x(u,v), y(u,v)\right)\right| = \left|\begin{array}{cc}
                            \frac{\partial x(u,v)}{\partial u} & \frac{\partial x(u,v)}{\partial v}\\
                            \frac{\partial y(u,v)}{\partial u} & \frac{\partial y(u,v)}{\partial v}
                            \end{array}
                            \right| = \left|\begin{array}{cc}
                            v & u\\
                            -v & 1-u
                            \end{array}
                            \right|=v.
\end{align}
Since the transformation is invertible, applying the change-of-variable formula yields
\begin{align}
f_{U,V}(u,v)&=\left|J\left(x(u,v), y(u,v)\right)\right|f_{X,Y}(x,y)\\
&=\frac{v}{\lambda_s^2}e^{-\frac{v}{\lambda_s}}.
\end{align}
Hence, we see obviously that random variables $U$ and $V$ are independent, and the random variable $U$ follows the uniform distribution with p.d.f.
\begin{align}
f_U(u) = 1, {0\leq U \leq 1},
\end{align}
and the random variable $V$ follows the gamma distribution with p.d.f.
\begin{align}
f_V(v) = \frac{v}{\lambda_{s}^2}e^{-\frac{v}{\lambda_{s}}}.
\end{align}
To this end, let $W=V|g|^2$ and $T=\frac{U}{|\hat{f}|^2}$, it is not difficult to show that the c.d.f. of $W$ and $T$ can be obtained as
\begin{align}\label{eqn:cdfW}
F_W(w)=1-2\frac{w}{\lambda_{s}\lambda_{d}}K_2\left(2\sqrt{\frac{w}{\lambda_{s}\lambda_{d}}}\right),
\end{align}
and
\begin{align}\label{eqn:cdfT}
F_T(t) = \lambda_{r}t\left(1-e^{-\frac{1}{\lambda_{r}t}}\right),
\end{align}
respectively.

Having characterized the statistical property of $W$ and $T$, we are ready to study the outage probability of the system, which is computed by
\begin{align}
P_{\sf out}^{\sf AF} &={\sf Pr}\left(\gamma_{\sf AF}\leq \gamma_{\sf th}\right) \notag\\
&= {\sf Pr}\left(W\left(\frac{TP_s}{N_0d_1^md_2^m}-\frac{\gamma_{\sf th}kP_s}{N_0d_1^md_2^m}\right)\leq \frac{T\gamma_{\sf th}}{k}+\gamma_{\sf th}\right)\notag\\
&=\left\{\begin{array}{cc} {\sf Pr}\left(W\leq \frac{\frac{T\gamma_{\sf th}}{k}+\gamma_{\sf th}}{\frac{TP_s}{N_0d_1^md_2^m}-\frac{\gamma_{\sf th}kP_s}{N_0d_1^md_2^m}}\right),& T>{k\gamma_{\sf th}}\\
                1, & T<{k\gamma_{\sf th}}\\
                \end{array}
                \right..
\end{align}
Then, invoking the c.d.f. of $W$ and $T$, the desired result can be obtained after some algebraic manipulations.

\section{Proof of Proposition \ref{prop:8}}\label{app:prop:8}
The end-to-end SNR can be alternatively expressed as
\begin{align}
\gamma_{\sf DF} = \min\left(\frac{T}{k},\frac{kP_sW}{d_1^md_2^mN_0}\right),
\end{align}
where $T$ and $W$ are defined in the proof of Proposition \ref{theorem:af2}. Hence, the outage probability is given by
\begin{align}
P_{\sf out}^{\sf DF}&= {\sf Pr}\left( \min\left(\frac{T}{k},\frac{kP_sW}{d_1^md_2^mN_0}\right)\leq \gamma_{\sf th}\right)
\end{align}
Due to the independence of random variables $T$ and $W$, the outage probability is given by
\begin{align}
P_{\sf out}^{\sf DF}&=1-{\sf Pr}\left(T\geq k\gamma_{\sf th}\right){\sf Pr}\left(W\geq \frac{\gamma_{\sf th}N_0d_1^md_2^m}{kP_s}\right).
\end{align}
To this end, invoking the c.d.f. expressions given in (\ref{eqn:cdfW}) and (\ref{eqn:cdfT}) yields the desired result.

\section{Proof of Proposition \ref{prop:9}}\label{app:prop:9}
The ergodic capacity can be expressed as
\begin{align}
R_{\sf AF} = {\tt E}\left\{\log_2\left(1+\frac{{aWT}}{\frac{T}{k}+{kaW}+1}\right)\right\},
\end{align}
where $T$ and $W$ is defined in the proof of Proposition \ref{theorem:af2}, $a =\frac{P_s}{N_0d_1^md_2^m}$. Alternatively, it can be written as
\begin{align}
&R_{\sf AF}
={\tt E}\left\{\log_2\left(1+\frac{T}{k}\right)\right\}+{\tt E}\left\{\log_2\left(1+kaW\right)\right\}\notag\\
&-{\tt E}\left\{\log_2\left(1+\frac{T}{k}+kaW\right)\right\}\label{eqn:caf1}\\
&={\tt E}\left\{\log_2\left(1+kaW\right)\right\}-{\tt E}\left\{\log_2\left(1+\frac{ka}{1+T/k}W\right)\right\}.
\end{align}

We now look at the first item, which can be alternatively evaluated via
\begin{align}
&{\tt E}\left\{\log_2\left(1+kaW\right)\right\} = \frac{1}{\ln 2}\int_0^{\infty}\frac{ka(1-F_W(w))}{1+kaw}dw\notag\\
&=\frac{2ka}{\lambda_{s}\lambda_{d}\ln 2}\int_0^{\infty}\frac{wK_2\left(2\sqrt{\frac{w}{\lambda_{s}\lambda_{d}}}\right)}{1+kaw}dw.
\end{align}
Then, utilizing the following identity \cite[pp. 54]{Meijer}
\begin{align}
\frac{x^{t}}{1+c x^{k}}=c^{-\frac{t}{k}}G_{1,1}^{1,1}\left(\left.cx^{k}\right|_{\frac{t}{k}}^{\frac{t}{k}}\right),
\end{align}
we have
\begin{multline}
{\tt E}\left\{\log_2\left(1+kaW\right)\right\}\\ =\frac{2}{\lambda_{s}\lambda_{d}\ln 2}\int_0^{\infty}G_{1,1}^{1,1}\left(\left.kaw\right|_{1}^{1}\right)K_2\left(2\sqrt{\frac{w}{\lambda_{s}\lambda_{d}}}\right)dw.
\end{multline}
Applying the integral identity \cite[Eq. (7.821.3)]{Table}, the integral can be solved as
\begin{align}\label{eqn:333}
{\tt E}\left\{\log_2\left(1+kaW\right)\right\}& =\frac{1}{\ln 2}G_{3,1}^{1,3}\left(\left.ka\lambda_s\lambda_d\right|_{1}^{-1,1,1}\right).
\end{align}
Finally, with the help of (\ref{eqn:333}), the second item can be expressed in integral form as
\begin{align}
{\tt E}\left\{1+\frac{ka}{1+T/k}W\right\} &=\frac{1}{\ln 2}\int_0^{\infty}G_{3,1}^{1,3}\left(\left.\frac{ka\lambda_s\lambda_d}{1+t/k}\right|_{1}^{-1,1,1}\right)\times\notag\\
&\left(\lambda_{r}-\left(\lambda_{r}+\frac{1}{t}\right)e^{-\frac{1}{\lambda_{r}t}}\right)dt.
\end{align}
To this end, pulling everything together yields the desired result.

\section{Proof of Corollary \ref{coro:1}}\label{app:coro:1}
Starting from (\ref{eqn:caf1}), we first look at the first item, which can be computed by
\begin{align}
&{\tt E}\left\{\log_2\left(1+\frac{U}{kZ}\right)\right\} = \frac{1}{\lambda_{r}}\int_0^{\infty}e^{-\frac{z}{\lambda_{r}}}dz\int_0^1\log_2\left(1+\frac{u}{kz}\right)du\\
&=\frac{1}{\lambda_{r}\ln 2}\int_0^{\infty}\left(-1+\left(1+kz\right)\ln\left(1+\frac{1}{kz}\right)\right)e^{-\frac{z}{\lambda_{r}}}dz\\
&=-\frac{1}{\ln 2}+\frac{1}{\lambda_{r}\ln 2}\int_0^{\infty}\left(1+kz\right)\left(\ln\left(1+{kz}\right)-\ln kz\right)e^{-\frac{z}{\lambda_{r}}}dz.
\end{align}
Then, with the help of the following identity
\begin{align}
\int_0^{\infty}\ln(1+x)x^{n-1}e^{-v x}dx=\frac{\Gamma(n)e^{v}}{v^n}\sum_{k=1}^{n}E_k(v),
\end{align}
we can compute
\begin{multline}
\int_0^{\infty}(1+kz)\ln(1+kz)e^{-\frac{z}{\lambda_r}}dz = \lambda_re^{\frac{1}{k\lambda_r}}E_1\left(\frac{1}{k\lambda_r}\right)+\\k\lambda_r^2e^{\frac{1}{k\lambda_r}}\sum_{k=1}^2E_k\left(\frac{1}{k\lambda_r}\right).
\end{multline}
Similarly, with the help of the identity \cite[Eq. (4.352.1)]{Table},
we can compute
\begin{multline}
\int_0^{\infty}(1+kz)\ln kze^{-\frac{z}{\lambda_r}}dz= (\lambda_r+k\lambda_r^2)\ln k+\lambda_r(\psi(1)+\ln \lambda_r)\notag\\
+k\lambda_r^2(\psi(2)+\ln\lambda_r).
\end{multline}
Hence, we have
\begin{multline}\label{eqn:222}
{\tt E}\left\{\log_2\left(1+\frac{U}{kZ}\right)\right\} =-\frac{1}{\ln 2}+\\
\frac{e^{\frac{1}{k\lambda_r}}}{\ln 2}\left(E_1\left(\frac{1}{k\lambda_r}\right)+k\lambda_r\sum_{k=1}^2E_k\left(\frac{1}{k\lambda_r}\right)
\right)\\
-\frac{1}{\ln 2}\left(\psi(1)+\ln k\lambda_r+k\lambda_r(\psi(2)+\ln k \lambda_r)\right).
\end{multline}
Since the second item has been evaluated in (\ref{eqn:333}), we now bound the third item. Again, using the fact that $f(x,y)=\log_2(1+e^x+e^y)$ is a convex function with respect to $x$ and $y$, the third item can be lower bounded by
\begin{multline}
{\tt E}\left\{\log_2\left(1+\frac{T}{k}+kaW\right)\right\} \\\geq \log_2\left(1+\frac{1}{k\lambda_r}e^{-1-\psi(1)}+ka\lambda_s\lambda_de^{2\psi(1)}\right).
\end{multline}
To this end, pulling everything together yields the desired result.


\nocite{*}
\bibliographystyle{IEEE}
\begin{footnotesize}

\end{footnotesize}

\end{document}